\documentclass{article}
\usepackage[hyphens]{url} 
\usepackage{smc2025}
\usepackage[caption=false, font=footnotesize]{subfig}
\usepackage{paralist}
\usepackage[figure,table]{hypcap}

\usepackage{booktabs}
\usepackage{multirow}
\usepackage{nicefrac}

\usepackage[english]{babel}
\usepackage[T1]{fontenc}

\def\papertitle{Music Boomerang: Reusing Diffusion Models\\for Data Augmentation and Audio Manipulation}

\author[1]{\mbox{\firstname{Alexander}\lastname{Fichtinger}}}
\author[1]{\mbox{\firstname{Jan}\lastname{Schlüter}\orcid{0000-0003-3862-6888}}}
\author[1,2]{\mbox{\firstname{Gerhard}\lastname{Widmer}\orcid{0000-0003-3531-1282}}}

\affil[1]{\department{Institute of Computational Perception}\institution{Johannes Kepler University Linz}\country{Austria}\affiliationtype{University}}
\affil[2]{\department{LIT AI Lab}\institution{Linz Institute of Technology}\country{Austria}\affiliationtype{University}}

\completesetup

\title{\papertitle}

\begin{document}
\capstartfalse
\maketitle
\capstarttrue

\begin{abstract}
Generative models of music audio are typically used to generate output based solely on a text prompt or melody. Boomerang sampling, recently proposed for the image domain, allows generating output close to an existing example, using any pretrained diffusion model.
In this work, we explore its application in the audio domain as a tool for data augmentation or content manipulation. 
Specifically, implementing Boomerang sampling for Stable Audio Open, we augment training data for a state-of-the-art beat tracker, and attempt to replace musical instruments in recordings.
Our results show that the rhythmic structure of existing examples is mostly preserved, that it improves performance of the beat tracker, but only in scenarios of limited training data, and that it can accomplish text-based instrument replacement on monophonic inputs.
We publish our implementation to invite experiments on data augmentation in other tasks and explore further applications.
\end{abstract}

\section{Introduction}\label{sec:introduction}
Generative models are designed to learn the underlying probability distribution of data, then generate new samples that match the characteristics of the training dataset. In the field of music generation, these models have to capture complex distributions that define elements such as melody, rhythm and harmony. Diffusion models \cite{ho2020denoising} in particular set the current state of the art. These models use iterative denoising processes to transform random noise into structured audio, enabling applications such as music generation \cite{evans2024long}, speech synthesis \cite{popov2021gradtts}, and sound design \cite{liu2023_audioldm}.

\begin{figure}
    \centering
    \subfloat[Diffusion]{\label{fig:diff}\includegraphics[scale=.4,trim=0 0 0 12]{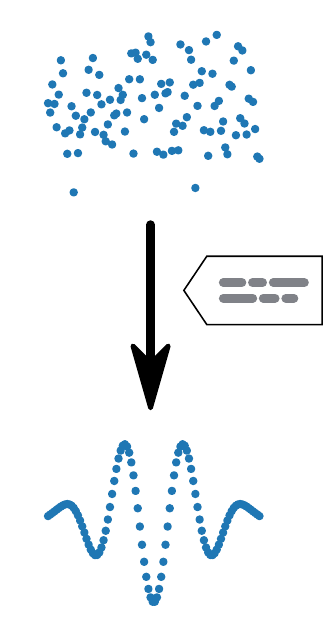}}
    \quad
    \subfloat[Boomerang sampling]{\label{fig:boom}\includegraphics[scale=.4,trim=0 0 0 12]{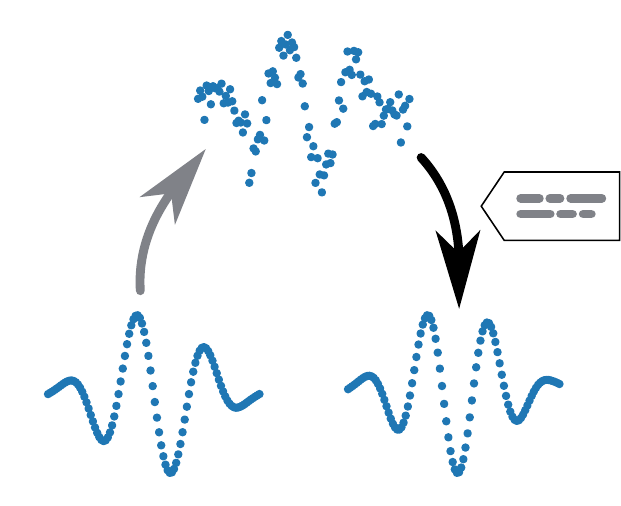}}
    \caption{(a) Diffusion models can gradually transform pure noise into audio. (b) Starting from a slightly noisy real example results in a variation of the example. In both cases, sampling may be controlled by text.}
    \label{fig:1}
\end{figure}

For most applications, generation is primed with pure noise (Fig.\ \ref{fig:diff}),
referred to as global sampling. Recent work has explored alternative sampling strategies to improve efficiency or to enable a more controlled generation. Among them, the Boomerang method \cite{luzi2024boomerang}, originally introduced for image generation, has demonstrated a way to perform local sampling by partially adding noise to an input sample and then employing a pretrained diffusion model to reverse this process, ending up close to the original (Fig.\ \ref{fig:boom}). This technique allows for subtle transformations while aiming to preserve key features of the input.

In this work, we adapt Boomerang sampling to the audio domain using a pretrained diffusion model. We explore two scenarios: 1) creating uncontrolled slight variations in music while maintaining important features such as rhythm and tempo, and 2) controlled manipulation of content. For the first one, we evaluate whether Boomerang sampling is suitable as a data augmentation method for machine learning tasks, comparing it to existing augmentation methods in beat and downbeat tracking. For the second scenario, we provide examples for musical timbre transformation.

The next section summarizes relevant prior work. Sec.~\ref{sec:boomerang_sampling} introduces Boomerang sampling, and Sec.~\ref{sec:implementation} details our adoption for the audio domain. Sec.~\ref{sec:experiments} describes our experimental results, and Sec.~\ref{sec:conclusions} concludes the paper.

\section{Related Work}
\label{sec:related_work}
For context and background, we will give a brief overview on modern strands of generative audio models and on controlled sampling from diffusion models.

\subsection{Audio Generation using Deep Learning}
Deep learning has significantly advanced audio generation, with models evolving from early autoregressive approaches to modern diffusion-based techniques. A popular method, WaveNet~\cite{oord2016wavenet}, generates new audio samples in a fully autoregressive manner. It uses a deep convolutional network (CNN) to model the conditional distribution of raw audio waveforms, employing causal dilated convolutions. Another method, called Jukebox~\cite{dhariwal2020jukebox}, proposes a hierarchical approach that uses vector quantized variational autoencoders (VQ-VAEs) to compress the raw audio into a lower-dimensional space. It generates new samples by learning a prior distribution over quantized encodings. 

Unlike autoregressive models that generate elements sequentially, non-autoregressive models produce entire sequences simultaneously. Within this category, Musika~\cite{marco_pasini_2022_7316720} utilizes a stacked autoencoder to encode log-magnitude spectrograms and employs a generative adversarial network (GAN) to model sequences of encoded vectors. MAGNeT~\cite{ziv2024masked}, another non-autoregressive technique, uses a masked generative approach, employing EnCodec~\cite{fossez2023high} as its latent representation and a single transformer to predict a subset of latent token spans, conditioned on previously generated past and future tokens.

Diffusion models~\cite{ho2020denoising} are particular relevant to this work. One of them is Stable Audio~\cite{pmlr-v235-evans24a}. It compresses audio into an invertible latent representation and uses a U-Net-based~\cite{ronneberger2015u} diffusion architecture inspired by Moûsai \cite{schneider-etal-2024-mousai}. Stable Audio Open~\cite{evans2024stable} follows a similar approach, but uses different model architectures and is trained only with audio files under a Creative Commons license.

\subsection{Advanced Sampling for Diffusion Models}
To improve efficiency and control in sampling, alternative strategies to global sampling (Fig.\ \ref{fig:diff}) have been developed.
A central idea among these approaches is the notion of starting the reverse diffusion process from an intermediate noisy signal rather than from pure Gaussian noise. To our knowledge, this concept was first introduced in SDEdit~\cite{meng2022sdedit}, which applied partial forward diffusion to an image before reconstructing it in the reverse process to create and edit photorealistic images from sketches. The key insight behind SDEdit was that introducing moderate noise (typically in the range of $t_0 \in [0.3, 0.6]$) destroys high-frequency details while preserving global structure, allowing meaningful modifications while maintaining fidelity, e.g., balancing realism and faithfulness.

Subsequent works have extended this principle beyond image editing. Chung et al. \cite{Chung2021ComeCloserDiffuseFasterAC} used forward-diffused initialization for inverse problems, demonstrating that starting from a more informed initialization can significantly accelerate the reverse diffusion process. Their framework highlights that lower levels of noise ($t_0 \leq 0.2$) are often sufficient when combined with auxiliary neural networks to refine the signal, improving efficiency without sacrificing quality. Similarly, Zheng et al. \cite{zheng2023truncated} investigated partial reverse diffusion by leveraging an intermediate noisy data distribution learned through a separate generative model, thus decreasing the number of necessary denoising steps. Finally, consistency models \cite{song2023consistencymodels} pushed this idea further by training networks to directly map noisy representations to clean samples, effectively collapsing the iterative reverse process into a few inference steps.

Boomerang sampling \cite{luzi2024boomerang} can be understood as part of this broader family of intermediate-step diffusion techniques. Unlike SDEdit, which was designed for conditional image editing/generation based on input sketches, Boomerang focuses explicitly on local sampling, generating slight variations of an input image while preserving its core characteristics. In contrast to methods such as those by Zheng et al.\ \cite{zheng2023truncated} and Chung et al.\ \cite{Chung2021ComeCloserDiffuseFasterAC}, Boomerang does not introduce an auxiliary generative model or require network modifications -- rather, it applies the standard reverse process starting from an intermediate timestep. 
Boomerang is particularly attractive for data augmentation, and has been used for image augmentation by Luzi et al. \cite{luzi2024boomerang}.
	
\section{Boomerang Sampling}\label{sec:boomerang_sampling}
Boomerang sampling~\cite{luzi2024boomerang} is a local sampling technique originally introduced for image generation with pretrained diffusion models. 
The underlying principle transfers naturally to the audio domain, and in this work we explore its application to music audio recordings.

The core idea is to apply a suitable noise level $n_\text{Boom}$ for perturbing a given sample $\textbf{x}_0$, allowing the model to 
introduce local refinements while aiming to retain important aspects of the original sample. This is controlled by a hyperparameter $t_\text{Boom}$, which defines the depth of the forward diffusion process starting from $\textbf{x}_0$. Afterwards, the reverse diffusion process reconstructs a denoised version $\textbf{x}_0'$ from $t = t_\text{Boom}$ to $t = 0$.

\subsection{Forward Process}
Considering an input sample $\textbf{x}_0$, the first step is to apply a forward diffusion process, described in closed form by
$$
    q(\textbf{x}_{t_\text{Boom}} \mid \textbf{x}_0) = \mathcal{N}(\sqrt{\bar{\alpha}_{t_\text{Boom}}} \textbf{x}_0, (1-\bar{\alpha}_{t_\text{Boom}}) \textbf{I}),
$$
where $\bar{\alpha}_t = \prod_{i=1}^t \alpha_i$ represents the accumulated noise schedule. Sampling from $q$, we obtain $\textbf{x}_{t_\text{Boom}}$.

\subsection{Reverse Process}
From $\textbf{x}_{t_\text{Boom}}$, the diffusion model 
reconstructs the signal through a reverse diffusion process, effectively denoising the sample to obtain a refined version $\textbf{x}_0'$. 
The reverse process follows standard stochastic  diffusion-based sampling, where each step is modelled as a conditional Gaussian:
$$
    p_\theta(\textbf{x}_{t-1} \mid \textbf{x}_{t}) = \mathcal N(\mu_\theta(\textbf{x}_{t},t), \sigma_t^2 \textbf{I})
$$
where $\mu_\theta$ represents the learned mean function, and $\sigma_t$ defines the noise schedule of the reverse process.

This process must be carried out step by step, exactly as in global sampling of diffusion models, and this is where most of the computational effort occurs. However, Boomerang sampling incurs significantly lower computational cost compared to global sampling. Since it only requires denoising from a partially noised sample, the cost is reduced by approximately \nicefrac{$t_\text{Boom}$}{$T$} compared to full diffusion, where $T$ is the total number of denoising steps. For instance, if $t_\text{Boom} = 0.2T$, sampling is 5 times faster than global sampling.

State-of-the-art implementations of diffusion models often use different mathematical representations and solvers to improve efficiency and flexibility. The forward and reverse processes can be expressed as stochastic differential equations (SDEs) or as ordinary differential equations (ODEs)~\cite{song2021scorebased, meng2022sdedit}. Advanced solvers \cite{dpm-solver-fast-ode, lu2023dpmsolverplusplus} optimize these formulations for faster and more accurate sampling.




\section{Music Boomerang} \label{sec:implementation}
To apply Boomerang sampling to audio generation, we integrate it with the publicly available Stable Audio Open model~\cite{evans2024stable}.
In the following, we first provide a brief overview of the model, explaining its key components and functionality, then explain in detail how this model is used in the Boomerang sampling pipeline.

\subsection{Stable Audio Open}
Stable Audio Open\cite{evans2024stable} is a latent diffusion model designed to generate stereo audio up to 47 seconds in length at a sample rate of 44.1\,kHz, based on text-prompts. It consists of three components: an autoencoder that compresses raw audio waveforms into a latent embedding, a T5-based text encoder~\cite{t5-encoder} for text conditioning, and a transformer-based diffusion model~\cite{evans2024long} that operates within the autoencoder's latent space.

\subsubsection{Autoencoder}
The authors used a variational autoencoder that operates on raw audio waveforms, utilizing an encoder with five convolutional blocks for downsampling and channel expansion through strided convolutions. ResNet-like layers with dilated convolutions and Snake~\cite{snake} activations are applied before each block to enhance processing. In the central layer of the autoencoder, the audio information is compressed into a compact representation with a latent dimension of 64. The decoder mirrors the encoders structure by using transposed convolutions for upsampling and reducing the number of feature channels. All convolutions are weight-normalized, and the decoders' output omits the $tanh$ activation to prevent harmonic distortion.

\subsubsection{Diffusion-Transformer (DiT)}
Stable Audio Open utilizes a diffusion-transformer (DiT)~\cite{evans2024long}, using stacked blocks of serially connected attention layers and gated multi-layer perceptrons (MLPs) with skip connections. An additional bias-less normalization was applied at the input to both the attention and MLP layers. Linear transformations are used at the input and output of the transformer to encode the latent dimension to the transformer's embedding dimension. Rotary positional embeddings (RoPE) \cite{RoPE} are applied to the key and query embeddings to make attention scores position-sensitive, where they are only applied to half of the embedding dimensions. Cross-attention layers are included in each block to integrate conditioning information. The transformer is guided by three conditioning signals: a text prompt, a timing input for handling variable-length generation and a timestep input, indicating the current step in the diffusion process.

\subsubsection{Training \& Inference}
The training procedure of the autoencoder involves three objectives. The primary objective is a reconstruction loss based on a perceptually weighted multi-resolution short-time fourier transform (STFT) for stereo audio, considering both mid-side (M/S) and left-right (L/R) channel representations. The L/R representation is down-weighted by a factor of 0.5 to address ambiguity around audio placement. In addition to the reconstruction loss, a second objective involves an adversarial loss with feature matching, using 5 convolutional discriminators similar to EnCodec~\cite{fossez2023high}. To ensure that the latent space of the autoencoder follows a desired distribution, a KL-divergence term is also employed.

The diffusion-transformer is trained using the v-objective method~\cite{salimans2022progressive} to predict noise increments from noised ground-truth latents. Specifically, the v-objective predicts a velocity vector $v \equiv \alpha_t \epsilon - \sigma_t \textbf{x}_0$, which gives $\hat{\textbf{x}}_0 = \alpha_t \textbf{x}_t - \sigma_t \hat{\textbf{v}_\theta}(\textbf{x}_t)$. The model uses DPM-Solver++ \cite{lu2023dpmsolverplusplus} with 100 decoding steps for efficient sampling.

\subsection{Boomerang Pipeline}
The Boomerang sampling pipeline is implemented as a modification to the Stable Audio Open inference process, enabling partial resampling. The pipeline consists of three primary phases: 1) Encoding the audio into the latent representation, 2) applying partial forward diffusion, and 3) performing the partial denoising process. Each of these steps is carefully adapted to ensure compatibility with the latent diffusion model and its scheduler functions.

\subsubsection{Encoding Input Audio}
Before applying Boomerang, the input audio is preprocessed and encoded into the model's latent space. The waveform is resampled (if necessary), peak-normalized, and padded/truncated to fit the model's expected input length. The raw waveform $\textbf{x}_0$ is passed through the Stable Audio Open autoencoder, producing a latent representation:
$$
    \textbf{z}_0 = E(\textbf{x}_0)
$$
where $E(.)$ is the encoder function. This latent representation serves as the starting point for the diffusion process.

\subsubsection{Partial Forward Diffusion}
Instead of fully diffusing the latent representation to pure noise, we want to specify a noise level $n_\text{Boom} \in (0,1)$. As the model's reverse diffusion noise schedule is not linear, we search the schedule for the time step $t_\text{Boom}$ that is closest to the desired noise level. From this, we compute the noisy latent in a single step:
$$
    \textbf{z}_{t_\text{Boom}} = \textbf{z}_0 + \sigma_{t_\text{Boom}} \epsilon, \quad \epsilon \sim \mathcal{N}(\textbf{0}, \textbf{I})
$$ 
with $\sigma_t$ controlling the unnormalized noise magnitude.

\subsubsection{Partial Reverse Diffusion}
The noisy latents are denoised using the diffusion model's reverse process, guided by a text prompt and a negative prompt. 
The denoising process begins at $t_\text{Boom}$, reconstructing the original latent representation while introducing local variations. For global sampling, Stable Audio Open uses $T=100$ steps. For Boomerang sampling, the number of steps is $t_\text{Boom} < 100$, which depends on the selected noise level.

Finally, the reconstructed latent representation $\textbf{z}_0'$ is transformed back into a raw audio waveform through the decoder $D(.)$ of the autoencoder:
$$
    \textbf{x}_0' = D(\textbf{z}_0').
$$


\subsubsection{Infinite-Length Generation via Frozen Overlaps}
\begin{figure}
    \centering
    \subfloat[First window]{\shortstack{\includegraphics[scale=.5]{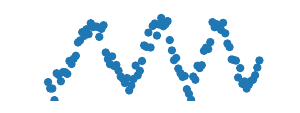}\\
    \includegraphics[scale=.4]{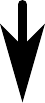}\\
    \includegraphics[scale=.5]{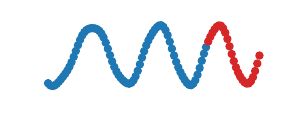}}}
    \subfloat[Second window]{\shortstack{\includegraphics[scale=.5]{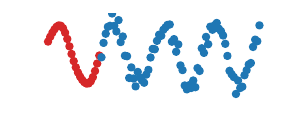}\\
    \includegraphics[scale=.4]{img/arrow.pdf}\\
    \includegraphics[scale=.5]{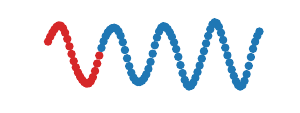}}}
    \caption{(a) The first window of a longer piece is sampled unconstrained. (b) To ensure coherence on boundaries, the second window overlaps the first one by 25\%, holding the overlapped portion fixed throughout all diffusion steps.}
    \label{fig:overlap}
\end{figure}

Stable Audio Open is limited to generating audio snippets of a fixed length (47 seconds). To apply Boomerang sampling to longer inputs while maintaining temporal coherence and avoiding boundary artifacts, we process it sequentially in 47-second windows overlapping by 25\%.
Specifically, after passing an excerpt $\textbf{x}_0^{(i)}$ through the encoder $E(.)$ to obtain the latent representation $\textbf{z}_0^{(i)}$ and applying the forward diffusion process up to $\textbf{z}_{t_\text{Boom}}^{(i)}$, we replace its first 25\% with the last 25\% of the previously generated $\textbf{z}_0'^{(i-1)}$ (i.e., the part where the windows overlap). This part is held constant (``frozen'') throughout the reverse diffusion process, enabling the model to generate a seamless continuation. Fig.~\ref{fig:overlap} illustrates this process, albeit on short waveforms instead of 47-second latent tensors.

\section{Experiments} \label{sec:experiments}
We conduct three experiments with Boomerang sampling on music recordings via Stable Audio Open, investigating whether it
1) preserves the rhythmic structure of a recording,
2) is suitable as data augmentation for beat tracking, and
3) enables prompt-based content manipulation.


\subsection{Analysis of Rhythmic Structure Preservation}
Our first goal is to determine how well Boomerang sampling preserves the rhythmic structure across different noise levels, a prerequisite for using it for data augmentation in the context of beat tracking.
We evaluate this with quantitative measures and qualitative observations.

\subsubsection{Quantitative Experiments}
For the quantitative evaluation, we process a set of music recordings with Boomerang sampling and evaluate whether onsets and beats remain detectable in the same positions.
This should provide a measure of the preservation of low-level musical structure after transformation (dependent on the effectiveness of the detection algorithms).
Specifically, onsets and beats are detected in both the original and transformed audio using librosa~\cite{mcfee2015librosa}, and then compared in terms of F1-score using mir\_eval \cite{raffel2014mir_eval}, treating detections in the original and transformed input as ground truth and predictions, respectively.
Evaluation uses a tolerance window of 80\,ms within which detected onsets or beats are considered to match.
The evaluation is conducted on the GTZAN genre dataset \cite{gtzan}. A generic text prompt (``Music'') and negative prompt (``Low Quality'') is used for all generations, with a guidance scale of 1.0 to prioritize the audio input over the prompt.

\begin{table}[t]
  \centering
    \begin{tabular}{lcll}
        \toprule
            \multicolumn{1}{c}{GTZAN genre} & \multicolumn{1}{c}{$n_\text{Boom}[\%]$} & \multicolumn{1}{c}{Onset F1} & \multicolumn{1}{c}{Beat F1} \\
        \midrule
            \multirow{4}{*}{\textit{all}} & $20$ & $87.2 \pm 5.7$ & $87.9 \pm 15.6$ \\
            & $40$ & $83.2 \pm 6.8$ & $82.0 \pm 18.8$ \\
            & $60$ & $78.3 \pm 8.3$ & $77.4 \pm 20.9$ \\
            & $80$ & $66.9 \pm 11.9$ & $56.0 \pm 24.1$ \\
            \cmidrule(lr){1-4}
            \textit{reggae} & $40$ & $89.0 \pm 4.7$ & $93.2 \pm 15.0$ \\
            \textit{disco} & $40$ & $88.5 \pm 4.9$ & $95.8 \pm 9.2$ \\
            \textit{pop} & $40$ & $88.0 \pm 6.7$ & $87.3 \pm 18.5$ \\
            \textit{hiphop} & $40$ & $87.3 \pm 4.4$ & $85.9 \pm 19.1$ \\
            \textit{rock} & $40$ & $82.8 \pm 7.7$ & $82.7 \pm 19.9$ \\
            \textit{metal} & $40$ & $82.7 \pm 5.4$ & $86.0 \pm 20.8$ \\
            \textit{blues} & $40$ & $81.7 \pm 6.6$ & $81.7 \pm 18.3$ \\
            \textit{country} & $40$ & $81.6 \pm 7.2$ & $86.9 \pm 17.4$ \\
            \textit{jazz} & $40$ & $78.8 \pm 8.3$ & $66.6 \pm 22.6$\\            
            \textit{classical} & $40$ & $71.3 \pm 12.2$ & $54.2 \pm 27.3$ \\
        \bottomrule
    \end{tabular}
    \caption{
    Evaluation of onset and beat preservation after Boomerang sampling with different noise levels on the GTZAN genre dataset. 
    We report the means and standard deviations over the music pieces.}
    \label{tab:grav1}
\end{table}

Our results show that both onsets and beats are partly preserved, but less and less reliably when increasing the noise level (Table~\ref{tab:grav1}, top part).
This suggests that higher noise levels distort the signal to the point where Boomerang sampling does not reproduce the original rhythmic and structural elements.
Comparing results by genre (Table~\ref{tab:grav1}, bottom part), we find that Reggae, Disco and Pop achieve the highest preservation scores. These genres often have well-defined rhythmic patterns, which may contribute to their robustness. Jazz and Classical fare worst. These are the most difficult genres for automatic beat trackers, so small input variations can already derail them.

\subsubsection{Qualitative Observations}

Additionally, we make several qualitative observations to capture perceptual differences while testing different noise levels (audio examples to follow along are available at \href{https://malex1106.github.io/boomify/}{malex1106.github.io/boomify/}). At low noise levels, the transformed audio closely resembles the original, with rhythmic patterns and harmonic structures. As the noise level increases to moderate levels, minor artifacts appear, but the rhythmic structure remains largely intact. High noise levels often lead to significant changes in the sound, which sometimes makes the original unrecognizable. This result depends heavily on the music genre. Still, even at high noise levels, Boomerang sampling maintains the overall structure of the piece. For all genres, we observe consistent difficulties in reconstructing vocal elements, a known limitation of Stable Audio Open \cite[p.\,4]{evans2024stable}.

\subsection{Beat Tracking Data Augmentation}
Beat tracking, the task of estimating beat positions in an audio signal, is one of the most well-known problems in the field of music information retrieval (MIR). The state of the art addresses it with machine learning on a large annotated dataset further expanded with pitch shifting and time stretching \cite{foscarin2024beatthis}. However, changing pitch or tempo does not capture all the potential variability in music. Our findings indicate that at low noise levels, Boomerang sampling preserves rhythmic structure while introducing small variations. Motivated by this, we explore Boomerang sampling as a data augmentation technique for beat tracking, aiming to enhance model generalization beyond the dimensions captured by pitch shifting and time stretching.

\subsubsection{Hyperparameter Search}

While musical variability increases with the noise level of Boomerang sampling, it also reduces the preservation of beats (Table~\ref{tab:grav1}).
In our first set of data augmentation experiments, we evaluate which noise level balances this tradeoff. We base our experiments on the published code of Foscarin et~al. \cite{foscarin2024beatthis}, disabling their data augmentation. For better reproducibility, we only use four datasets with openly available audio (ballroom, candombe, guitarset and tapcorrect) instead of all fifteen datasets used by Foscarin et~al., and increase the number of training epochs from 100 to 400 to counter the reduced dataset size. We precompute two variations of each file using Boomerang sampling with $n_\text{Boom}$ of 20, 40, 60 or 80\%. During training, in each epoch we pick randomly among the three versions (the original and its two variations). We evaluate beat and downbeat detection performance in terms of F1-score, again using mir\_eval \cite{raffel2014mir_eval}, comparing against human annotations.
Our results on the validation split (Table~\ref{tab:hyper}, top part) show that a noise level of 40\% performs best among the four levels tested. Surprisingly, it is also the only one that improves beat detection performance over not augmenting at all.

\begin{table}
    \centering
        \begin{tabular}{ccll}
            \toprule
            $n_\text{Boom}[\%]$ & Variations & \multicolumn{1}{c}{Beat F1} & \multicolumn{1}{c}{Downbeat F1} \\
            \midrule
            -- & 0 & $95.1 \pm 0.4$ & $89.8 \pm 1.3$ \\
            \cmidrule(lr){1-4}
            20 & 2 & $94.4 \pm 0.2$ & $89.2 \pm 1.4$ \\
            \textbf{40} & 2 & $\textbf{95.4} \pm \textbf{0.3}$ & $\textbf{90.2} \pm \textbf{1.7}$ \\
            60 & 2 & $94.9 \pm 0.3$ & $90.2 \pm 1.2$ \\
            80 & 2 & $94.4 \pm 0.6$ & $88.4 \pm 0.8$ \\
            \cmidrule(lr){1-4}
            40 & 1 & $95.0 \pm 0.2$ & $88.6 \pm 1.2$ \\
            40 & 2 & $95.4 \pm 0.3$ & $90.2 \pm 1.7$ \\
            40 & 4 & $95.4 \pm 0.1$ & $90.3 \pm 0.4$ \\
            40 & \textbf{6} & $\textbf{95.6} \pm \textbf{0.3}$ & $\textbf{90.4} \pm \textbf{0.5}$ \\
            40 & 8 & $95.1 \pm 0.7$ & $90.1 \pm 0.3$ \\
            \bottomrule
        \end{tabular}
    \caption{Hyperparameter search on a train/validation split of the reduced dataset (ballroom, candombe, guitarset and tapcorrect). We report means and standard deviations over 3 runs. Optimal parameters marked in bold.}
    \label{tab:hyper}
\end{table}

In a second set of experiments, we vary the number of variations per file. Increasing it captures more variability in music, but reduce the exposure to the original file (as we pick randomly among all available versions), which is the only one that the annotations can be fully trusted for. Our results (Table~\ref{tab:hyper}, bottom part) indicate that 6 variations are optimal, although most settings are close.

\subsubsection{Augmentation Comparison}

To assess the effectiveness of Boomerang sampling compared to existing augmentation methods or no augmentation at all, we evaluate models on the GTZAN dataset \cite{gtzan}, an established test set for beat tracking. We use Boomerang sampling with the optimal hyperparameters of the previous section (noise level of 40\%, 6 variations per file), and compare four variants: Training without any augmentation, training with Boomerang sampling, training with the default time stretching, pitch shifting and masking augmentations of Foscarin et~al. \cite{foscarin2024beatthis}, and training with combining the default augmentations with Boomerang sampling.


Training on the train and validation split of the reduced dataset (1181 files) from the hyperparameter search, both Boomerang sampling and the default augmentations result in slight gains for beat detection and stronger ones for downbeat detection (Table~\ref{tab:eval_gtzan_small}). Boomerang sampling alone is less effective than the default augmentations, but improves F-scores by about 3 pp.\ when combined with them.

\begin{table}
    \centering
        \begin{tabular}{lcc}
            \toprule
            Augmentation & \multicolumn{1}{c}{Beat F1} & \multicolumn{1}{c}{Downbeat F1} \\
            \midrule
            No augmentation & $79.5\pm 1.7$ & $57.1\pm 0.7$ \\
            Boomerang sampling & $79.7 \pm 0.3$ & $60.2\pm 0.6$ \\
            Pitch, tempo, masks & $79.8\pm 1.1$ & $64.1\pm 0.7$ \\
            \textbf{All augmentations} & $\textbf{82.5}\pm \textbf{1.7}$ & $\textbf{67.8}\pm \textbf{1.7}$ \\
            \bottomrule
        \end{tabular}
    \caption{Evaluation on the test dataset (GTZAN) after training on the train and validation split of the reduced dataset using the optimal parameters from Table~\ref{tab:hyper}. We report means and standard deviations over 3 runs.}
    \label{tab:eval_gtzan_small}
\end{table}

Training on the full dataset (4556 files) used in Foscarin et~al.\ \cite{foscarin2024beatthis}, we obtain qualitatively similar results (Table~\ref{tab:eval_gtzan}):
Augmentations improve downbeat detection more strongly than beat detection, Boomerang sampling alone is less effective than the default augmentations, but improves results when combined.
However, as results are generally closer together, the improvement over the state of the art through Boomerang sampling is marginal. We conclude that it is mainly useful in regimes of small training sets.

\begin{table}
    \centering
        \begin{tabular}{lcc}
            \toprule
            Augmentation & \multicolumn{1}{c}{Beat F1} & \multicolumn{1}{c}{Downbeat F1} \\
            \midrule
            No augmentation & $88.2\pm 0.1$ & $75.0\pm 0.9$ \\
            Boomerang sampling & $88.5 \pm 0.1$ & $76.2\pm 0.6$ \\
            Pitch, tempo, masks \cite{foscarin2024beatthis} & $89.1\pm 0.3$ & $78.3\pm 0.4$ \\
            \textbf{All augmentations} & $\textbf{89.2}\pm \textbf{0.3}$ & $\textbf{78.6}\pm \textbf{0.2}$ \\
            \bottomrule
        \end{tabular}
    \caption{Evaluation on the test dataset (GTZAN) after training on the train and validation split of the full dataset. We report means and standard deviations over 3 runs. Results for pitch, tempo, masks augmentation are taken from Foscarin et al.~\cite{foscarin2024beatthis}.}
    \label{tab:eval_gtzan}
\end{table}

\subsection{Content Manipulation}
In addition to evaluating the uncontrolled generation of minor variations with Boomerang sampling, we explore its potential for targeted audio manipulation and style transfer. By adjusting the text prompt, guidance scale and noise level, it is possible to introduce controlled modifications (again, audio samples to follow along are available at \href{https://malex1106.github.io/boomify/}{malex1106.github.io/boomify/}).

In particular, solo instruments can be modified by changing the text prompt while keeping the original musical structure close to the original sample. For example, using a guitar sample as input and applying the text prompt ``Trumpet lead'' with higher noise levels and guidance scales can modify the timbre while preserving the underlying harmonic framework.
At lower noise levels, the transformation sometimes blends the source and target timbre, while higher noise levels produce more pronounced changes that may not only replace the instrument but also cause deviations from the original melody.

Polyphonic recordings can be altered as well, but as the text prompt controls all of the generation in the reverse diffusion process, it will be applied to all components of the mix -- it is not possible to ask for changing a particular voice. For example, applying the prompt ``Trumpet solo'' to a pop song will modify both vocals and background instruments to resemble trumpets.

Overall, these findings demonstrate the potential of Boomerang sampling for content modification, allowing changes in musical attributes without completely discarding the original musical structure. However, the effectiveness of such transformations depends on the complexity of the original input, the given text prompt, the guidance scale and the level of noise introduced during the process.

\section{Conclusions} \label{sec:conclusions}
In this work, we adapted Boomerang sampling \cite{luzi2024boomerang} to the audio domain by implementing it for a pretrained diffusion model, Stable Audio Open \cite{evans2024stable}. Our results show that it is suitable for data augmentation in training a beat and downbeat detector, but more so if only limited training data is available. We also show that by guiding the sampling process with a text prompt, we can use it to manipulate existing audio material in a controlled way.

In the context of data augmentation, two directions would be interesting to explore: 1) Instead of using the neutral prompt ``Music'', a set of different prompts could promote a larger variability of augmented versions, and 2) instead of merely hoping that the sampling process keeps the annotated qualities -- in our case, the beat and downbeat positions -- intact, we could guide the sampling process by the existing annotations. However, the latter requires retraining or finetuning the diffusion model to be conditioned on the annotations first.

At \href{https://github.com/malex1106/boomify}{https://github.com/malex1106/boomify} we publish audio samples and code to reproduce our experiments or try Boomerang sampling online, hoping to invite further research or artistic applications.

\pagebreak  

\begin{acknowledgments}
This work was supported by the European Research Council (ERC) under the EU’s Horizon 2020 research \& innovation programme, grant agreement No.\ 101019375 (Whither Music?), and the Federal State of Upper Austria (LIT AI Lab).
\end{acknowledgments} 

\bibliography{smc2025bib}
	
\end{document}